\documentclass[
onecolumn,
superscriptaddress,
reprint,
 amsmath,amssymb,
 aps, physrev,
]{revtex4-2}

\usepackage{graphicx}
\usepackage{dcolumn}
\usepackage{bm}
\usepackage{xfrac}
\usepackage{comment}
\usepackage{mathtools}
\usepackage{physics}
\usepackage{makecell}

\newcommand{\Ham}{\hat{H}}

\newcommand{\Id}{\hat{I}}
\newcommand{\dm}{\hat{\rho}}

\newcommand{\Lin}{\mathcal{L}}

\begin{document}

\preprint{APS/123-QED}

\title{\textbf{From the Bloch equation to the thermodynamically consistent master equation}
}

\author{Eugenia Pyurbeeva}
\email{eugenia.pyurbeeva@mail.huji.ac.il}
\affiliation{The Institute of Chemistry and the Fritz Haber Center for Theoretical Chemistry, The Hebrew University of Jerusalem, Jerusalem 9190401, Israel}
\author{Ronnie Kosloff}
\affiliation{The Institute of Chemistry and the Fritz Haber Center for Theoretical Chemistry, The Hebrew University of Jerusalem, Jerusalem 9190401, Israel}

\date{\today}
\begin{abstract}
The Bloch equation that set the foundation for open quantum systems, was conceived by pure physical reasoning. Since then, the Lindblad (GKLS) form of a quantum master equation, its most general mathematical representation, became an established staple in the open quantum systems toolbox. It allows to describe a multitude of quantum phenomena, however its universality comes at a cost -- without additional constraints, the resultant dynamics are not necessarily thermodynamically consistent, and the equation itself lacks an intuitive interpretation. 
\\
We present a mathematically equivalent form of the Lindblad master equation under a single constraint of strict energy conservation. The ``elemental Bloch'' equation separates the system dynamics into its elemental parts, making an explicit distinction between thermal mixing, dephasing, and energy relaxation, and thus reinstating the physical intuition in the equation. We derive the equation for a many-level system by accounting for all 
relevant transitions between pairs of levels.
Finally, the formalism is illustrated by calculating the fixed point of the dynamics and exploring the conditions for canonical invariance in quantum systems. 
\end{abstract}
\maketitle
\section{\label{sec:intro}Introduction}
The original form of the quantum master equation was a guess on the part of Felix Bloch, in an attempt to describe the dynamics of nuclear polarisation vector coupled to a thermal bath \cite{bloch1946nuclear}. Bloch combined the unitary dynamics of the polarisation vector with its relaxation to its equilibrium parameters, stressing the analogy
with Heisenberg equation of motion, and defining two separate relaxation timescales: $T_1$ for energy relaxation, and $T_2$
for the phase, or the components of the polarisation perpendicular to the constant magnetic field.

To establish the theory on more solid ground, Bloch \cite{wangsness1953dynamical,bloch1956dynamical},
Fano \cite{fano1954note,fano1957description}, and Redfield \cite{redfield1957theory,redfield1965advances} rederived the master equation from time-dependent perturbation theory. The overarching idea was to start from a global Hamiltonian
$\Ham_G$, partition it into the system, environment and interaction
Hamiltonians:
$$\Ham_G =\Ham_S+\Ham_E + \Ham_{SE}$$
and then treat the interaction Hamiltonian $\Ham_{SE}$ as a small expansion parameter.

The Bloch equation, originally developed for NMR, has evolved far beyond its initial purpose and has become a cornerstone of the theory of open quantum systems.
Numerous derivations were proposed, based
on perturbation theory \cite{laird1991quantum,lampert2025sixth},
projection operator techniques \cite{nakajima1958quantum,zwanzig1960ensemble,gonzalez2024tutorial}, or the collision model \cite{karplus1948note,dumcke1985low}.
E.B. Davis presented a mathematically more rigorous
derivation of the master equation with $\hat H_{SE}$ as the small parameter \cite{davies1974markovian}, which leads to a thermal fixed point of the dynamics.

A major paradigm change in the field was the introduction of a mathematical template for the most general form of the master equation,
known as the Gorini, Kossakowski, Lindblad and Sudarshan (GKLS) form \cite{lindblad1976generators,gorini1976completely}:
\begin{equation}
  \frac{\dd}{\dd t}{{\dm_S} }=   -\frac{i}{\hbar}[{\Ham_S};{ \dm_S }] +\Lin \left({\dm_S}\right),
  \label{eq:lindblad}
\end{equation}
with the dissipative part 
$\Lin(\dm_S)$ having the  structure: 
 \begin{equation}
     \Lin (\dm_S)=\sum_{j}\gamma_{j}
    \left( \hat{L}_{j} \dm_{S}\hat{L}_{j}^{\dagger} \\-\frac{1}{2}\{\hat{L}_{j}^{\dagger}\hat{L}_{j} ;\dm_{S} \}\right) 
    \label{eq:dissipator}
 \end{equation}
where $\hat L_j$ are known as the Lindblad jump operators, which are not determined by the framework, and the kinetic coefficients $\gamma_{j}$ are real and positive. The GKLS template has had a profound influence on quantum open-system theory \cite{breuer2002theory}, allowing to construct
models of diverse phenomena, from quantum optics \cite{allen2012optical} and quantum information processing \cite{rivas2012open} to
quantum computing \cite{kashyap2025accuracy}. 

Nevertheless, this mathematical formulation lost Bloch's direct connection to physical reality. In particular, the fixed point of the dynamics is not necessarily a thermal state. In order to obtain a thermodynamically consistent master equation, additional constraints need to be imposed, notably, strict energy conservation $[\Ham_{SE},\Ham_S +\Ham_E]=0$ \cite{dann2021open,alicki1976detailed}. 
This restriction is equivalent to the secular
approximation \cite{trushechkin2021unified,jung2025quantum}.

In this work, we explore the GKLS master equation under the imposition of strict energy conservation constraint and find a mathematically equivalent expression that brings back the intuitiveness of the original Bloch master equation form, allowing to separate the energy relaxation, level population mixing, and thermal dephasing terms. 

We start with a simple case of a two-level system, and then extend our considerations to a system with arbitrarily many levels by decomposing the evolution into a sum of two-level system dynamics. Finally, we demonstrate the utility of our master equation by exploring the conditions for canonical invariance.

\section{The main idea \label{sec:main}}

A major consequence of the strict energy conservation constraint is that the dissipative and the unitary parts of Eq. (\ref{eq:lindblad}) commute, and thus have common set of eigenoperators \cite{dann2021open}. 

The eigenoperator set of the unitary part $[\Ham_S; \bullet]$ consists of projection operators $\hat P_i=\ket{i}\bra{i}$, invariant to the dynamics, and transition operators $\hat F_{ij}=\ket{j}\bra{i}$, which exist in pairs ($\hat F_{ij} =\hat F_{ji}^\dagger$) and form the set of Lindblad jump operators.

However, the GKLS equation leaves the overall scaling of the Lindblad jump operators free, as the transformation $\hat{L}_j \rightarrow \alpha \hat{L}_j$, $\gamma_j \rightarrow \gamma_j/(\alpha \alpha^*)$ preserves the dynamics. 

We will impose an additional constraint on the jump operator scaling, ensuring that consequent up and down jumps leave the state intact, instead of multiplied by a constant. A choice of scaling does not further narrow the scope of dynamics described by the GKLS equation with an energy conservation constraint imposed, and simply reduces a linear space of mathematically equivalent equations to a single point by convention. 

To explore the implications of our choice, we begin by analysing a two-level system and deriving a master equation under our imposed restrictions. A more general master equation for a system with arbitrarily many levels will be derived following on the two-level system analysis, as a decomposition of many-level dynamics into that of all pairs of interacting energy eigenstates. 

\section{A two-level system \label{sec:two}}
\subsection{The Hamiltonian and jump operators}
For a two-level system, we define an almost-arbitrary Hamiltonian:
\begin{equation}
    \Ham=\frac{E}{2}\left(\varepsilon_z \sigma_z +\varepsilon_x \sigma_x+\varepsilon_y \sigma_y \right),
\end{equation}
where $\sigma_i$'s are the Pauli matrices, with a constraint of $\varepsilon_z^2+\varepsilon_x^2+\varepsilon_y^2=1$ (from this point onwards, we will omit the subscript $S$ for system Hamiltonian and density matrix, as the GKLS equation \ref{eq:lindblad} does not refer to any others). In this definition, we have neglected the possible overall offset in energy, as it does not affect system dynamics. The two eigenstates then have energies of $\pm E/2$, with $E$ being the energy difference between the energy levels.

It is easy to show that if $\ket{s_0}$ and $\ket{s_1}$ are eigenstates of $\Ham$ with eigenvalues $-E/2$ and $E/2$ respectively, then $\ket{s_1}\bra{s_0}$ and $\ket{s_0}\bra{s_1}$ are eigenoperators of $[\Ham; \bullet]$ with eigenvalues $\pm E$:
\begin{multline}
    [\Ham; \ket{s_1}\bra{s_0} ]=\Ham \ket{s_1}\bra{s_0}-\ket{s_1}\bra{s_0}\Ham=\\=\frac{E}{2}\ket{s_1}\bra{s_0}-\ket{s_1}\bra{s_0}\left(-\frac{E}{2}\right)=E\ket{s_1}\bra{s_0}
\end{multline}
(and similarly for $\ket{s_0}\bra{s_1}$).

Since we wish to leave the freedom for the scaling of the jump operators, we define two jump operators $\sigma_p$ and $\sigma_m$, $\sigma_p=\sigma_m^\dagger$, eigenoperators of $[\Ham_S; \bullet]$ with eigenvalues $\pm E$, but so far merely proportional to $\ket{s_1}\bra{s_0}$ and $\ket{s_0}\bra{s_1}$. The other two eigenoperators of $[\Ham_S; \bullet]$ are $\Ham$ itself and $\Id$, both with zero eigenvalues.

\subsection{Jump operator algebra}
We now explore the multiplicative properties of $\sigma_p$ and $\sigma_m$. Trivially, $(\sigma_p \sigma_m)^\dagger=\sigma_m^\dagger \sigma_p^\dagger=\sigma_p \sigma_m$. The same for $\sigma_m \sigma_p$, hence both products are self-adjoint.  

\subsubsection{Jump operator squares}
As the first order of business, we show that $\sigma_p^2=\sigma_m^2=0$ by comparing $E\sigma^2_p$, which can be expressed as both $[\Ham; \sigma_p] \sigma_p$ and $\sigma_p [\Ham; \sigma_p]$, with $E^2\sigma^2_p$ -- the square of the commutator. From the former:
 \begin{multline}
    E \sigma_p^2=[\Ham; \sigma_p] \sigma_p=\sigma_p [\Ham; \sigma_p] \Leftrightarrow \\\Ham \sigma_p^2-\sigma_p \Ham \sigma_p=\sigma_p \Ham \sigma_p - \sigma_p^2 \Ham
\end{multline}
we obtain:
\begin{equation}
\sigma_p \Ham \sigma_p=\frac{1}{2}\{ \Ham; \sigma_p^2 \}
\end{equation}
Then, using this result and the fact that for our definition of the Hamiltonian, $\Ham^2=\frac{E^2}{4}\Id$, we proceed to the latter:
\begin{multline}
    E^2 \sigma_p^2 = [\Ham; \sigma_p] [\Ham; \sigma_p]=\left(\Ham \sigma_p-\sigma_p \Ham \right)^2=\\ =\Ham (\sigma_p \Ham \sigma_p)-\frac{E^2}{4}\sigma_p^2-\Ham \sigma_p^2 \Ham + (\sigma_p \Ham \sigma_p) \Ham=\\
    = \frac{\Ham}{2} \{ \Ham; \sigma_p^2 \}-\frac{E^2}{4}\sigma_p^2-\Ham \sigma_p^2 \Ham +\{ \Ham; \sigma_p^2 \}\frac{\Ham}{2}=\\=\frac{E^2 \sigma_p^2}{8}+\frac{\Ham \sigma_p^2 \Ham}{2} -\frac{E^2 \sigma_p^2}{4} -\Ham \sigma_p^2 \Ham +\frac{\Ham \sigma_p^2 \Ham}{2}+\frac{E^2 \sigma_p^2}{8} =0
\end{multline}
The same holds true for $\sigma_m^2$. 

\subsubsection{Commutator}
To find the commutator $[\sigma_p; \sigma_m]$, we write the cyclic relation:
\begin{equation}
    [\Ham;[\sigma_p; \sigma_m]]+[\sigma_p;[\sigma_m; \Ham]]+[\sigma_m;[\Ham; \sigma_p]]=0.
\end{equation}
Together with the commutation relations of the jump operators, this leads to:
\begin{multline}
    [\Ham;[\sigma_p; \sigma_m]]+E[\sigma_p;\sigma_m]+E[\sigma_m;\sigma_p]=\\=[\Ham;[\sigma_p; \sigma_m]]=0
\end{multline}
Since $[\sigma_p; \sigma_m]$ commutes with $\Ham$, it must belong to the linear operator space based on $\Id$ and $\Ham$, the eigenoperators of the free evolution of the system. Due to our choice of a traceless Hamiltonian and the trace of the commutator of any two operators being zero, the jump operator commutator $[\sigma_p; \sigma_m]$ must simply be proportional to $\Ham$, and due to the freedom of scaling of the jump operators in the GKLS equation, we are free to choose the coefficient of proportionality.

We can thus make a decision:
\begin{equation}
\label{eq:comm}
     [\sigma_p; \sigma_m]=\frac{2\Ham}{E}
\end{equation}

Using this scaling, and the squares of the jump operators being zero, as shown above, we can return to the commutation of jump operators with $\Ham$ as:
\begin{multline}
    [[\sigma_p; \sigma_m];\sigma_p]=(\sigma_p \sigma_m-\sigma_m \sigma_p)\sigma_p-\sigma_p(\sigma_p \sigma_m-\sigma_m \sigma_p)=\\=2\sigma_p \sigma_m \sigma_p=2\sigma_p
\end{multline}
(and similarly for commutation with $\sigma_m$), which leads to a pair of triple relations:
\begin{equation}
\label{eq:triple}
    \begin{dcases}
    \sigma_p \sigma_m \sigma_p=\sigma_p\\
    \sigma_m \sigma_p \sigma_m=\sigma_m
    \end{dcases}
\end{equation}

\subsubsection{Anticommutator}
The triple relations above (Eq.\ref{eq:triple}), combined with the squares $\sigma_p^2=\sigma_m^2=0$ allow to show that the anticommutator $\{\sigma_p; \sigma_m \}$ commutes with both $\sigma_p$ and $\sigma_m$ -- two non-commuting operators. Moreover:
\begin{equation}
    \begin{dcases}
        \{\sigma_p; \sigma_m \}\sigma_p=(\sigma_p\sigma_m+\sigma_m\sigma_p)\sigma_p=\sigma_p\sigma_m\sigma_p=\sigma_p
        \\[5pt]
        \{\sigma_p; \sigma_m \}\sigma_m=(\sigma_p\sigma_m+\sigma_m\sigma_p)\sigma_m=\sigma_m\sigma_p\sigma_m=\sigma_m
    \end{dcases}
\end{equation}
which allows us to conclude that:
\begin{equation}
\label{eq:anticomm}
    \{\sigma_p; \sigma_m\}=\Id
\end{equation}
This relation, combined with the squares of the jump operators being equal to zero, fully reflects that for fermionic creation and annihilation operators. It is a natural consequence of the levels of a two-level system being mutually exclusive states, and could have been taken as a starting point. 

\subsection{The modified dissipator form}
For a two-level system, the standard GLKS dissipator following (Eq.\ref{eq:dissipator}) takes the form:
\begin{equation}
    \Lin(\dm)=\gamma_p \Lin_p(\dm)+\gamma_m \Lin_m(\dm)
\end{equation}
where $\gamma_p/\gamma_m$ follows detailed balance, and $\Lin_{p/m}(\dm)$ dissipative parts corresponding to addition and subtraction (creation and annihilation) respectively:
\begin{equation}
    \begin{dcases}
        \Lin_p(\dm)=\sigma_p \dm \sigma_m -\frac{1}{2} \{\sigma_m \sigma_p; \dm \}\\
        \Lin_m(\dm)=\sigma_m \dm \sigma_p -\frac{1}{2} \{\sigma_p \sigma_m; \dm \}
    \end{dcases}
\end{equation}
In this section we prove that under our definition of the Hamiltonian and the scaling of the jump operators $\sigma_{p/m}$ conforming to the ``natural'' commutation relations (Eqs. \ref{eq:comm},\ref{eq:anticomm})
the dissipative parts of the GKLS equation can be expressed as: 
\begin{equation}
\label{eq:therm-diss}
    \Lin_{p/m}(\dm)=\frac{1}{2}\Id-\dm \pm \frac{\Ham}{E}+\frac{[ \Ham [ \Ham; \dm]]}{2E^2}
\end{equation}
In adherence to the example set by Bloch, this form for the dissipator has been a guess on our part, and we proceed to justify it below.

In order to reduce redundancy, we prove the equivalence of $\Lin_p(\dm)\pm \Lin_m(\dm)$ expressed through jump operators and in our suggested thermodynamic form (Eq.\ref{eq:therm-diss}).
\begin{equation}
    \begin{dcases}
         \sigma_p \dm \sigma_m-\sigma_m \dm \sigma_p=\frac{2\Ham}{E}- \{ \frac{\Ham}{E};\dm \}    \\[5pt]
        \sigma_p \dm \sigma_m+\sigma_m \dm \sigma_p=\frac{[\Ham; [\Ham; \dm]]}{E^2}-\dm+\Id
    \end{dcases}
\end{equation}
The above can be further simplified by introducing the traceless density matrix $\dm_0=\dm-\frac{1}{2}\Id$:
\begin{equation}
\label{eq:to_prove}
    \begin{dcases}
        \sigma_p \dm_0 \sigma_m-\sigma_m \dm_0 \sigma_p=- \{ \frac{\Ham}{E};\dm_0 \}    \\
        \sigma_p \dm_0 \sigma_m+\sigma_m \dm_0 \sigma_p=\frac{[\Ham; [\Ham; \dm_0]]}{E^2}-\dm_0
    \end{dcases}
\end{equation}
\\
In the next step towards the proof, we notice that the anticommutator of two traceless matrices is always proportional to identity. It is the immediate consequence of the anticommutator of any two Pauli matrices $i$ and $j$ being equal to $2\Id \delta_{ij}$, where $\delta_{ij}$ is the Kronecker delta.

In our case, $\dm_0$ is traceless by definition, and $\sigma_{p/m}$ due to being proportional to a commutator $\sigma_{p/m}\sim[\Ham; \sigma_{p/m}]$, which is always traceless in itself. We can thus define:
\begin{equation}
    \begin{dcases}
        \{\sigma_p; \dm_0 \}=A \Id\\
         \{\sigma_m; \dm_0 \}=A^* \Id
    \end{dcases}
\end{equation}
This allows to write expressions for the traceless density matrix $\dm_0$ ``sandwiched'' between two jump operators:
\begin{equation}
    \begin{dcases}
        \sigma_p \dm_0 \sigma_m = A^* \sigma_p -\sigma_p \sigma_m \dm_0=A \sigma_m - \dm_0 \sigma_p \sigma_m \\
        \sigma_m \dm_0 \sigma_p = A^* \sigma_p -\dm_0 \sigma_m \sigma_p =A \sigma_m -  \sigma_m \sigma_p \dm_0 \\
        \sigma_p \dm_0 \sigma_p=A \sigma_p\\
        \sigma_m \dm_0 \sigma_m=A^* \sigma_m
    \end{dcases}
\end{equation}
and leads to proving equations \ref{eq:to_prove}:
\begin{multline}
    \sigma_p \dm_0 \sigma_m-\sigma_m \dm_0 \sigma_p=\\=\frac{1}{2} \left( A^* \sigma_p +A \sigma_m - \sigma_p \sigma_m \dm_0 -\dm_0 \sigma_p \sigma_m   \right)-\\-\frac{1}{2} \left( A^* \sigma_p +A \sigma_m - \dm_0 \sigma_m \sigma_p  - \sigma_m \sigma_p \dm_0   \right)=\\
    = \frac{1}{2}\left( \left(\sigma_m \sigma_p- \sigma_p \sigma_m \right)\dm_0 + \dm_0\left(\sigma_m \sigma_p- \sigma_p \sigma_m \right) \right)=\\=-\frac{1}{2}\{[\sigma_p;\sigma_m]; \dm \}=-\{\frac{\Ham}{E};\dm_0 \} 
\end{multline}
and
\begin{multline}
    \sigma_p \dm_0 \sigma_m+\sigma_m \dm_0 \sigma_p=\\=\frac{1}{2} \left( A^* \sigma_p +A \sigma_m - \sigma_p \sigma_m \dm_0 -\rho_0 \sigma_p \sigma_m   \right)+\\+\frac{1}{2} \left( A^* \sigma_p +A \sigma_m - \dm_0 \sigma_m \sigma_p  - \sigma_m \sigma_p \dm_0   \right)=\\
    = A^*\sigma_p +A \sigma_m -\\- \frac{1}{2}\left(\sigma_p \sigma_m \dm_0 +\dm_0 \sigma_p \sigma_m +\dm_0 \sigma_m \sigma_p +\sigma_m \sigma_p \dm_0   \right)=\\=A^*\sigma_p +A \sigma_m-\dm_0    
\end{multline}
together with:
\begin{multline}
     \frac{[\Ham; [\Ham; \dm_0]]}{E^2}=\frac{1}{2}\dm_0 -\frac{1}{2}\left( \frac{2\Ham}{E}\dm_0 \frac{2\Ham}{E}\right)=\\=\frac{1}{2}\left(\{\sigma_p; \sigma_m\}\dm_0\{\sigma_p; \sigma_m\}-[\sigma_p; \sigma_m]\dm_0 [\sigma_p; \sigma_m] \right)=\\
    =\frac{1}{2} \left(2 \sigma_m \sigma_p \dm_0 \sigma_p \sigma_m + 2 \sigma_p \sigma_m \dm_0 \sigma_m \sigma_p \right)=\\=\sigma_m A\sigma_p \sigma_m +\sigma_p A^* \sigma_m \sigma_p=A\sigma_m+A^*\sigma_p
\end{multline}
give us the desired result, and prove the validity of our proposed dissipator form (Eq.\ref{eq:therm-diss}).

\subsection{The full equation and stationary state}
Using the dissipator expression proved above (Eq.\ref{eq:therm-diss}), we can write the full master equation for a two-level system:
\begin{multline}
\label{eq:therm-rate-two}
    \frac{\dd \dm}{\dd t}=-i [ \Ham; \dm ]-\left(\gamma_p+\gamma_m\right)\left(\dm -\frac{1}{2}\Id \right)+\\+\left(\gamma_p-\gamma_m\right)\frac{\Ham}{E}+\left(\gamma_p+\gamma_m\right)\frac{[ \Ham [ \Ham; \dm]]}{2E^2}
\end{multline}
This form, which we coin the \textit{elemental Bloch equation} (EBE), due to it being reminiscent of the original Bloch equation, is equivalent to the general GKLS form with an energy conservation constraint. Its most prominent feature is that it partitions the dynamics into a sum of elemental terms, each with immediate physical interpretation.  

The first term remains the free evolution of the system, the second is the ``mixing'' term, driving the system to a state with equal level occupation probabilities, the third term is energy relaxation, and the last is thermal dephasing. To this term, we can add the third trivial eigenoperator of $\Ham$, generating the pure dephasing, $\Gamma [ \Ham [ \Ham; \dm]]$. As a result Eq. \ref{eq:therm-rate-two} unifies the description of thermalisation-induced dephasing, and pure dephasing $T_2$
and $T_2^*$ \cite{skinner1986pure,PhysRevLett.130.123601}.

In many cases, such as a fermionic bath or the scattering theory approach to a bosonic bath \cite{levy2012quantum}, the rates $\gamma_{p/m}$ can be written as:
\begin{equation}
    \begin{dcases}
        \gamma_p=\gamma f(E)\\
        \gamma_m=\gamma \left( 1-f(E)\right)
    \end{dcases}
\end{equation}
where $f(E)$ is the Fermi distribution with the temperature of the bath, and $\gamma$ is the coupling strength, which can be in itself temperature-dependent, but the detailed balance by the rates is satisfied by the Fermi distributions. Typically, the temperature dependence of $\gamma$ is weak, and both the mixing and dephasing rates in the EBE (Eq.\ref{eq:therm-rate-two}) can be thought of as temperature-independent. It is also worth noting that the dissipator form (Eq.\ref{eq:therm-diss}) shows that both occur in equal amounts for both directions of exchange with the bath.

As further demonstration of both validity and utility of our form of the GKLS equation, we find the stationary state for a two-level system from Eq. \ref{eq:therm-rate-two}. It can be done directly by assuming that the stationary state $\dm_S$ will commute with the system Hamiltonian $\Ham$. Then the master equation reduces to:
\begin{equation}
    \left(\gamma_p+\gamma_m\right)\left(\dm_S -\frac{1}{2}\Id \right)=\left(\gamma_p-\gamma_m\right)\frac{\Ham}{E}
\end{equation}
giving:
\begin{equation}
    \dm_S=\frac{1}{2}\Id + \frac{\left(\gamma_p-\gamma_m\right)}{\left(\gamma_p+\gamma_m\right)}\frac{\Ham}{E}
\end{equation}
which aligns with the Gibbs state:
\begin{equation}
    \dm_{\text{Gibbs}}=\frac{e^{-\frac{\Ham}{T}}}{\Tr \left(e^{-\frac{\Ham}{T}} \right)}
\end{equation}
as long as detailed balance $\gamma_p/\gamma_m=e^{-E/T}$ is satisfied.
\\
\section{Extending to arbitrarily many levels \label{sec:many}}
\subsection{The thermodynamic master equation}

Having found the EBE form of a quantum master equation for a two-level system, it is only reasonable to extend it to those with more complex dynamics and thus more energy levels. This extension, however, will not hold one of the greatest advantages of Eq. \ref{eq:therm-rate-two} -- the two-level form of the thermodynamic master equation does not require the Hamiltonian to be diagonal, it just calls upon the energy level spacing between the levels. However, if we are to extend a similar equation to higher dimensions, separate rate coefficients $\gamma_{p/m}$ for every possible transition will be necessary, and since the requirement for all energy levels to be known is unavoidable, the Hamiltonian may as well be considered diagonal. 

It is then natural to consider the evolution of each pair of energy levels exchanging populations as following an equation similar to Eq.\ref{eq:therm-rate-two}. In order to facilitate a mathematical description, we introduce a set of projection operators $\Id_t=\ket{s_i}\bra{s_i}+\ket{s_j}\bra{s_j}$
on the subspace of states $i$ and $j$, so that $\Ham_t=\Id_t\Ham \Id_t$ and $\dm_t=\Id_t\dm \Id_t$ -- the partial Hamiltonian and density matrix -- only include the 2x2 sub-operators responsible for the population exchange between the $i$-th and $j$-th levels, where $t$ is the index of exchange pairs without regard for the order.

Now, remaining in the spirit of Bloch, we endeavour to guess the form of the multi-level EBE. The major driving force on this path is the fact that the GKLS dissipator (Eq.\ref{eq:dissipator}) is linear in $\dm$, while our thermodynamic form (Eq.\ref{eq:therm-diss}) is not. To correct for it we propose:
\begin{multline}
\label{eq:therm-full}
    \frac{\dd \dm}{\dd t}=-i [\Ham; \dm]-\sum \limits_t(\gamma^t_p+\gamma^t_m)\left(\dm_t-\Tr \dm_t\frac{\Id_t}{2} \right)+\\+\sum \limits_t(\gamma^t_p-\gamma^t_m)\frac{\Ham_t-\Tr \Ham_t \dfrac{\Id_t}{2}}{E_t}\Tr \dm_t+\\+\sum \limits_t(\gamma^t_p+\gamma^t_m)\frac{[\Ham_t;[\Ham_t;\dm_t]]}{2E_t^2}
\end{multline}
where $\gamma^t_{p/m}$ are the exchange rates to and from the system between for the transition $t$, and $E_t$ -- the energy spacing between the levels in the transition. 

This proposal retains its intuitive interpretation: $\Tr\rho_t$ is the total population involved in the exchange, and it is  natural that the mixing term evolves the partial density matrix $\rho_t$ towards equal occupation of the two levels while retaining the total occupation probability. Similarly, in the energy relaxation term, each partial Hamiltonian has been deprived of its energy offset, which does not influence the inter-level dynamics in a single pair, and the proportionality of energy relaxation to the population of the levels involved is also in line with the expected behaviour. Finally, in the dephasing term, the energy offset does not need to be removed as the constant term in the Hamiltonian disappears in the double commutator. 
 
\subsection{The stationary state of a harmonic oscillator}
\label{sec:oscillator}

In order to justify our guess of the multi-level EBE form (Eq.\ref{eq:therm-full}) we use it to find a stationary state for a very specific case of a harmonic oscillator. 

We take the Hamiltonian to be:
\begin{equation}
    \Ham=\begin{pmatrix}
...&&&&&\\
&&3E&&\mbox{\LARGE 0}&\\
&&&2E&&\\
&\mbox{\LARGE 0}&&&E&\\
&&&&&0\\
\end{pmatrix}
\end{equation}
with infinitely many levels spaced by $E$, where transitions are only allowed between consecutive levels and the transition rates have the form:
\begin{equation}
    \begin{dcases}
       \gamma_p^t=\gamma_i f(E)\\
       \gamma_m^t=\gamma_i \left(1-f(E) \right)
    \end{dcases}
\end{equation}
where $t$ is the transition between the levels $i$ and $i+1$, $\gamma_i$ is the coupling strength, arbitrarily dependent on the transition level -- it thus represents a more general case than a typical harmonic oscillator where $\gamma_i=(i+1)\gamma$, where $\gamma$ is the geometric coupling strength, equal for all levels.

The master equation (Eq.\ref{eq:therm-full}) can then be simplified to:
\begin{multline}
\label{eq:osc-rate}
    \frac{\dd \dm}{\dd t}=-i [\Ham; \dm]- \sum \limits_{i}\gamma_i\left(\dm_i-\Tr \rho_i\frac{\Id_i}{2} \right)+\\+ \left(2f(E)-1\right) \sum \limits_{i}\gamma_i\frac{\Ham_i-\Tr \Ham_i \dfrac{\Id_i}{2}}{E}\Tr \dm_i+\\+ \sum \limits_{i}\gamma_i\frac{[\Ham_i;[\Ham_i;\dm_i]]}{2E^2}
\end{multline}
The traceless normalised partial Hamiltonians for pairs of consecutive levels will all be equal to $\sigma_z^i=1/2(\ket{s_{i+1}}\bra{s_{i+1}}-\ket{s_i}\bra{s_i})$, and since we are looking for a stationary state and assume it commutes with the Hamiltonian, the equation can be reduced to:
\begin{multline}
\label{eq:osc-eq}
    \sum \limits_{i}\gamma_i\left(\dm_i-\Tr \dm_i\frac{\Id_i}{2} \right)=\\=\left(2f(E)-1\right) \sum \limits_{i}\gamma_i\sigma_z^i\Tr \dm_i
\end{multline}
Taking the stationary density matrix to be diagonal, we can number its diagonal elements as $p_i$ -- the occupation probabilites of the $i$-th level -- and Eq.\ref{eq:osc-eq} becomes a series of linear equations on $p_i$. 

We prove that the stationary state for our oscillator is the Gibbs state in a classic two-step induction fashion.

\textbf{Basis:} For $p_0$, only involved in the exchange with $p_1$, Eq.\ref{eq:osc-eq} leads to:
\begin{equation}
    p_0-\frac{p_0+p_1}{2}=-\frac{1}{2}(2f(E)-1)\left(p_0+p_1 \right)
\end{equation}
which gives:
\begin{equation}
    p_1=e^{-\frac{E}{T}}p_0
\end{equation}

\textbf{Inductive step:} For $p_i$, involved in the exchange with both $p_{i+1}$ and $p_{i-1}$, Eq.\ref{eq:osc-eq} becomes:
\begin{multline}
    \gamma_i\left(p_i-\frac{\left(p_i+p_{i+1} \right)}{2}\right)+\gamma_{i-1}\left(p_i-\frac{\left(p_i+p_{i-1} \right)}{2}\right)=\\=\left(2f(E)-1\right)\left(-\frac{\gamma_i}{2}\left(p_i+p_{i+1} \right)+\frac{\gamma_{i-1}}{2}\left(p_i+p_{i-1} \right) \right)
\end{multline}
simplifying to:
\begin{multline}
\label{eq:step-simple}
    p_i \left(\gamma_i f(E)+\gamma_{i-1}\left(1-f(E) \right) \right)=\\=\gamma_{i-1}f(E)p_{i-1}+\gamma_i(1-f(E))p_{i+1}
\end{multline}
If $p_{i-1}=e^{\frac{E}{T}}p_i=p_i(1-f(E))/f(E)$, this further reduces to:
\begin{equation}
    p_{i+1}=\frac{f(E)}{\left(1-f(E)\right)}p_i=e^{-\frac{E}{T}}p_i
\end{equation}
thus completing the proof.

\section{Canonical invariance \label{sec:canon}}
As an application example of our form of the master equation, we use it to explore conditions for canonical invariance\cite{Andersen1964} -- asking the question of when the maximum entropy generalised Gibbs form is preserved during thermalisation dynamics.

For a two-level system, the generalised Gibbs state preservation statement is almost trivial: for any $T$, the Gibbs state: 
\begin{equation}
    \dm_G=\frac{e^{-\Ham/T}}{\Tr \left(e^{-\Ham/T} \right)}
\end{equation}
commutes with the Hamiltonian and thus belongs to the subspace of $\Id$ and $\Ham$ and can be written as $\dm_G=\Id/2+\lambda \Ham/E$.

Substituted into the master equation Eq.\ref{eq:therm-rate-two}, it gives:
\begin{equation}
    \frac{\dd \dm}{\dd t}=-\left(\gamma_p+\gamma_m\right)\lambda \frac{\Ham}{E}+\left(\gamma_p-\gamma_m\right)\frac{\Ham}{E}
\end{equation}
and as the derivative is proportional to $\Ham$, the state remains in the generalised Gibbs state, and thermalisation can be described as:
\begin{equation}
    \frac{\dd \lambda}{\dd t}=-\left(\gamma_p+\gamma_m\right)\lambda+\left(\gamma_p-\gamma_m\right)
\end{equation}

For a system with infinitely many equally spaced energy levels coupled to the thermal bath discussed in Section \ref{sec:oscillator}, a generalised Gibbs state requires $p_{i+1}/p_i=a$ for all values of $i$, where $a=\exp(-E/T)$ and $T$ evolves with time, giving:
\begin{equation}
    \frac{\dd }{\dd t}\left(\frac{p_{i+1}}{p_i} \right)=\frac{\dot{p_{i+1}}}{p_i}-\frac{a\dot{p_i}}{p_i}=\frac{\dd a}{\dd t}
\end{equation}
This condition simplifies to: 
\begin{equation}
\label{eq:Gibbs-crit}
   \frac{\dot{p_{i+1}}}{p_{i+1}}-\frac{\dot{p_i}}{p_i}=\frac{1}{a}\frac{\dd a}{\dd t}=\frac{\dd \ln a}{\dd t}
\end{equation}
which for a generalised Gibbs state has to be equal for all energy levels.

Similarly to Eq.\ref{eq:step-simple}, for a diagonal density matrix, the multi-level master equation (Eq.\ref{eq:therm-full}) gives:
\begin{multline}
\label{eq:dd-level}
    \frac{\dd p_i}{\dd t}=-\left(f(E)\gamma_i+\left(1-f(E) \right)\gamma_{i-1} \right)p_i+\\+\left(1-f(E) \right)\gamma_i p_{i+1}+f(E)\gamma_{i-1}p_{i-1}
\end{multline}
We introduce a parameter $\delta$ quantifying the difference between the momentary generalised Gibbs distribution of the system and the Gibbs distribution at the temperature of the bath:
\begin{equation}
    a=\frac{f(E)}{(1-f(E))}e^\delta
\end{equation}
where $f(E)$ is the Fermi distibution at the bath temperature.

Employing this new parameter, Eq.\ref{eq:dd-level} can be rewritten as:
\begin{equation}
\label{eq:dd-level-sim}
    \frac{1}{p_i}\frac{\dd p_i}{\dd t}=\gamma_i f(E)\left(e^\delta -1\right)+\gamma_{i-1}\left(1-f(E) \right)\left(e^{-\delta} -1\right)
\end{equation}
which allows us to write the criterion for the Gibbs form of the density matrix (Eq.\ref{eq:Gibbs-crit}) as:
\begin{multline}
\label{eq:crit-gen}
    \frac{\dot{p_{i+1}}}{p_{i+1}}-\frac{\dot{p_i}}{p_i}=\left(\gamma_{i+1}-\gamma_i \right)f(E)\left(e^\delta -1\right)+\\+\left(\gamma_i-\gamma_{i-1} \right)\left(1-f(E) \right)\left(e^{-\delta} -1\right)
\end{multline}
Equation \ref{eq:dd-level-sim} is for an energy level involved in the exchange with levels both above and below it. For $p_0$, the equivalent derivative and Gibbsianity criterion are:
\begin{equation}
    \frac{\dd p_0}{\dd t}=\gamma_0 f(E)\left(e^\delta-1 \right)p_0
\end{equation}
and
\begin{multline}
\label{eq:crit-0}
    \frac{\dot{p_1}}{p_1}-\frac{\dot{p_0}}{p_0}=\left(\gamma_1-\gamma_0 \right)f(E)\left(e^\delta-1 \right)+\\+\gamma_0 (1-f(E))\left(e^{-\delta}-1 \right)
\end{multline}
From the canonical criteria equations (Eq.\ref{eq:crit-gen} and Eq.\ref{eq:crit-0}), it follows that in order for the density matrix to remain in the generalised Gibbs state with time evolution, the coefficients $\gamma_i$ must satisfy:
\begin{equation}
    \gamma_{i+1}-\gamma_i=\gamma_0
\end{equation}
for every value of $i$. Which is precisely the case for a standard harmonic oscillator. 

Finally, from the above results, by restoring $a$, we can obtain the thermalisation equation:
\begin{equation}
\label{eq:therm}
    \frac{\dd \ln a}{\dd t}=e^{-\frac{E}{T(t)}}(1-f(E))+e^{\frac{E}{T(t)}}f(E)-1
\end{equation}
If $T=T_B$, the derivative in Eq.\ref{eq:therm} is equal to zero, as $\exp(-E/T)=f(E)/(1-f(E))$. It can also be seen that the derivative has the appropriate signs for thermalisation if $a\lessgtr\exp(-E/T_B)$.

\section{Discussion \label{sec:discussion}}

In this study, we introduced and demonstrated a novel form of a quantum master equation, which we name the \textit{elemental Bloch equation}. It is based on the GKLS equation -- the most general form of a quantum master equation -- with the added constraint of strict energy conservation. A ``natural'' scaling of the jump operators, maintains the scope of the dynamics described.  

The EBE form regains much of Bloch's direct physical interpretation of its terms -- it explicitly separates the dynamics of a system into a sum of elemental parts: free evolution, occupation mixing between energy levels, energy relaxation, and thermal and pure dephasing. An immediate consequence is that it rationalises the weak 
temperature dependence of thermal dephasing. 

Another advantage of the simple and intuitive form of the EBE is its facility for analytical work. We have demonstrated its application to finding the stationary state of a two-level system and a harmonic oscillator with arbitrary level couplings to the baths, as well as a study of the conditions for canonical invariance in these systems.

Finally, due to the equation containing the Hamiltonian and its contracted parts directly, with no need for jump operators to be found, the EBE may have an advantage over the GKLS form in computational methods, which will be explored in further work. 

\section*{Acknowledgements}
E.P. is grateful to the Azrieli Foundation for the award of an Azrieli Fellowship.



\end{document}